# Self-organized instability in graded-index multimode fibres

Logan G. Wright, Zhanwei Liu, Daniel A. Nolan, Ming-Jun Li, Demetrios N. Christodoulides, and Frank W. Wise

Multimode fibres (MMFs) are attracting interest for spatiotemporal dynamics, ultrafast fibre sources, imaging, and telecommunications. This interest stems from three differences compared to single-mode fibre (SMF) structures: their spatiotemporal complexity (information capacity), the role of disorder, and complex intermodal interactions. To date, MMFs have been studied in limiting cases where one or more of these properties can be neglected. Here we study a regime in which all these elements are integral. We observe a spatial beam-cleaning process preceding spatiotemporal modulation instability. We provide evidence that the origin of these processes is a universal unstable attractor in graded-index MMFs. The self-organization and instability of the attractor are both caused by intermodal interactions characterized by cooperating disorder, nonlinearity and dissipation. A disorder-enhanced nonlinear process in MMF has important implications for future telecommunications, and the multifaceted complexity of the dynamics showcases MM waveguides as potential laboratories for many topics in complexity science.

Multimode fibres are now at the core of many developments in optics. For imaging and telecommunications, MMFs offer unprecedented information density. For telecommunications, this facilitates increased bandwidth for internet traffic[1-7]. For imaging[8-10], it means a variety of high-resolution optical imaging modalities may be performed through a robust and flexible fibre endoscope with a diameter of ≈100-1000 μm. To access these features, researchers have tamed the spatiotemporal complexity of MMFs by complete measurement of the fibre's transmission (or transfer) matrix and its principal modes[4-10]. For spatiotemporally complex propagation in MMFs, these tools recover underlying order in the linear coupling between the modes, allowing spatial and temporal control of linear propagation.

Meanwhile, multimode fibres have also been studied in the nonlinear regime[11-35]. Nonlinear coupling between modes causes a wide range of novel effects, many of which may be useful for high-power ultrashort-pulsed fibre sources. A cubic nonlinearity couples up to four distinct waves. As a result, breaking down MM nonlinear dynamics into a mode-coupling picture requires 4-dimensional tensors[11], which describe the nonlinear coupling between the spatial eigenmodes. The dimension and nonlinearity restrict the amount of insight that can be directly inferred from the structure of these tensors compared to the linear transmission and group delay matrices, but they provide an efficient description that is useful in many situations.

This coupled-mode approach is not unique to optics. Today, research on many complex systems focuses primarily on the coupling of the systems' distinct elements. This is the key insight of complexity science: in many complex systems, important behaviours result from the features of the interactions between the distinct elements in the system. Of particular interest is the topology of the interactions: the complex network[36,37]. In contrast to optics, however, in many systems of interest in complexity science - including human social interactions, financial markets, and brains - experiments are not straightforward. In many cases, researchers rely instead on natural experiments or computer experiments. These difficulties have led to controversy (see, e.g. Refs. 38-39). For this reason, an experimental system like multimode fibre should be of broad scientific interest. It is highly controllable and measurable, and yet supports complex phenomena. It is essentially a large network of interacting dynamical systems. The topology of this network, and the nature of the connections, can be readily controlled by the mode structure of the fibre, the properties of the exciting field, the initial distribution of light within the modes, the presence of disorder or active media, and the length and orientation of the fibre.

Here we study self-organization of nonlinear waves in normal-dispersion, multimode graded-index (GRIN) fibres, in a regime where intermodal interactions are mediated by disorder, nonlinearity and dissipation. In this regime, we observe a spatial transformation of arbitrary input fields to a consistent attractor (*i.e.*, 'beam clean-up'), which is the fundamental mode with a weak background of higher-order modes. This attractor is unstable. Consequently, once it is reached we observe spatiotemporal modulation instability, which causes the field to evolve towards a spatiotemporally-complex steady-state. We provide a simplified theoretical model to understand the origin of this self-organized instability in terms of cooperative intermodal interactions.

Our study integrates a wide range of MM phenomena, and many of these individual physical processes have been studied to some extent before, including four wave mixing[15,16,18,20,24,25], stimulated Raman scattering (SRS)[12,13,32-34], and multiple nonlinear processes simultaneously[12,13,18,21-23,30,31,34]. Theoretical studies have explored the limit of strong fibre disorder in the context of nonlinear pulse propagation[26-29]. More broadly, many studies have explored strong disorder in optics[40-45], motivated in particular by Anderson localization. Research in this area has explored a variety of phenomena defined by disorder and dissipation; a key example is the random laser[45]. Nonlinear interactions have also been studied in disordered systems: they underlie a variety of interesting phenomena, as well as important open questions (see Refs. 41-44 and references therein). Compared to other systems with disorder, dissipation and nonlinearity, MMFs are valuable because these properties (as well as dispersion) can be easily and precisely adjusted over a broad range. Furthermore, because of the important applications mentioned above, understanding MMFs has clear value beyond basic science.

# Results

We launch nanosecond-duration pulses in the normal-dispersion regime of long, coiled GRIN multimode fibres, exciting a large number of transverse modes, and monitor the spatial and spectral profiles of the output field. Fig. 1 shows results from a typical experiment. The variation of spectral features (a) and spatial profile (b, c) of the output beam with increasing energy are shown for fixed 100-m fibre length. Fig. 1c shows the near-field beam profiles for increasing pulse energy. At low energy, the beam undergoes a dramatic transformation of self-focusing. Remarkably, this transformation is observed when imaging the full-spectrum beam (using achromatic lenses), rather than only a particular Stokes wave. By roughly 1.5 µJ, the field has reached a clean, Gaussian beam: the critical state of the attractor. With the attractor reached, the field is acutely unstable and spectral sidebands appear. These sidebands result from multiple four-wave mixing processes with spatiotemporal phase-matching[15].

Accurate modelling of these observations using standard field-envelope propagation equations is prohibitively difficult. Disorder, dissipation, and Raman and Kerr nonlinearities are all important, and the field occupies 55 modes with a time-bandwidth product that exceeds $2 \cdot 10^5$ (*i.e.,* [1 nanosecond]· [200 THz]). Many complex systems exhibit winner-take-all[37] and self-organized critical dynamics[46, 47], and complex network analysis has proven useful when more exact, complete models are impossibly time-consuming or so complicated that they fail to provide much insight. Hence, in what follows we will describe the multifaceted nature of the mode-coupling -- the complex network -- in the GRIN fibre in order to find a qualitative understanding of the experiments.

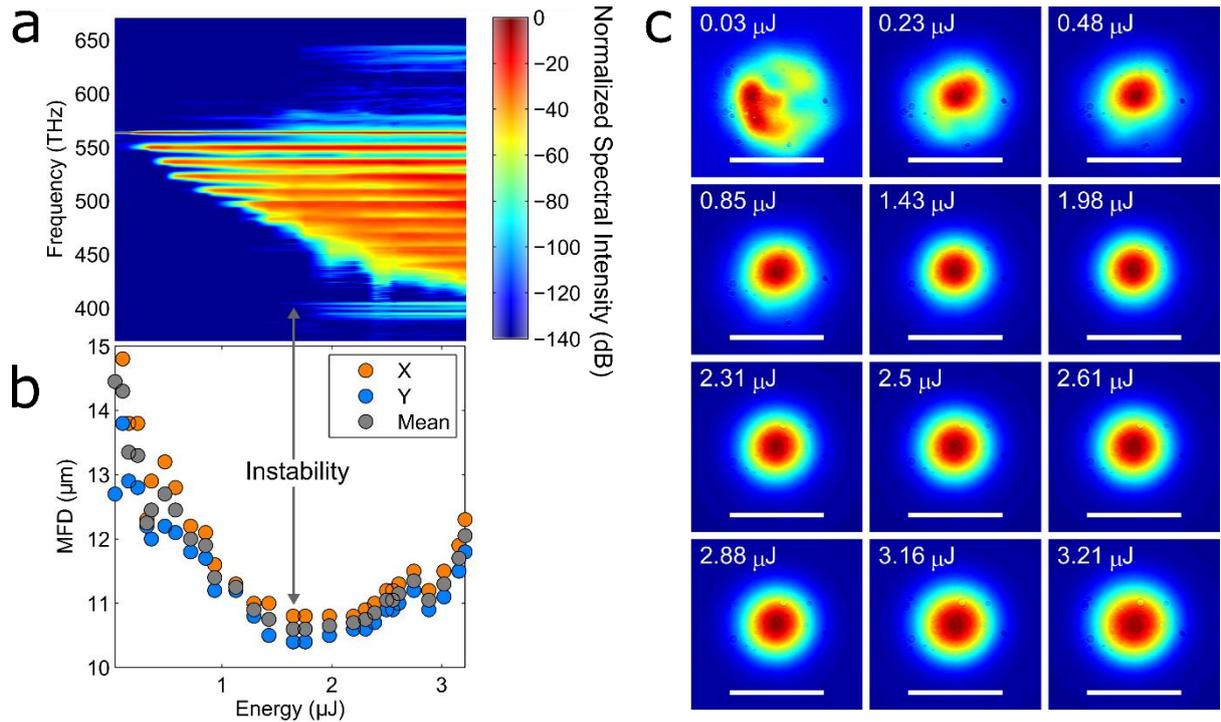

**Figure 1: Experimental measurements of self-organized instability in normal-dispersion GRIN fibre.** (a) The spectrum of the field exiting the fibre as a function of increasing launched energy. (b) The spatial size (mode field diameter, MFD) of the field over the same range. (c) Output field spatial profile for increasing energy launched into the fibre. As energy increases, the field organizes until it has reached the unstable attractor (about 1.5 μJ): the fundamental mode on a background of higher-order modes. Increasing the energy from this point leads to spatiotemporal modulation instability and spatiotemporal complexity. Scale bar is 11 μm. Evolution is shown in Supplementary Movie 1. For visibility, the spectrum at each energy is normalized to the peak spectral intensity at that energy.

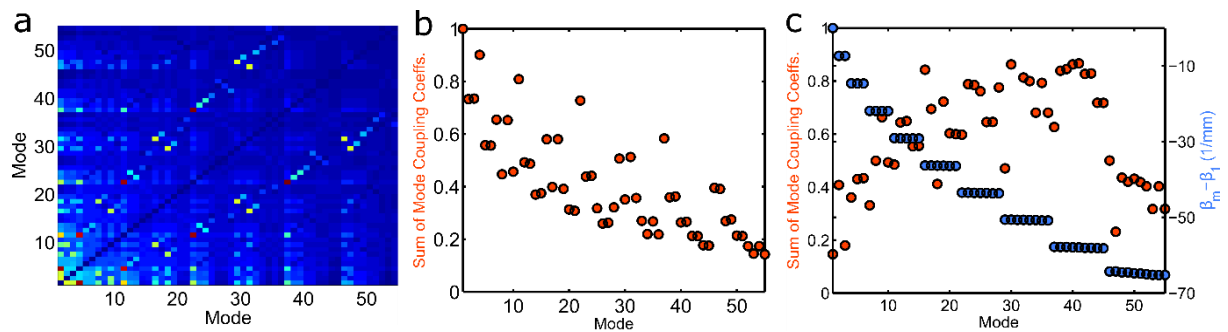

**Figure 2: Characteristics of mode-coupling in GRIN MMF.** (a) Magnitude of typical nonlinear mode-coupling matrix, at a particular fibre position, $z'$, $\overline{|C_{lm}(z')|}$. (b) Sum of the rows of the nonlinear mode-coupling matrix, $\sum_{m=1}^{55}|C_{lm}(z')|$. (c) Sum of the rows of a typical disorder-induced mode-coupling matrix (See Supplementary Figure 1-2), and the propagation constants (relative to the fundamental mode) of the modes. The decrease of the disorder-induced coupling for the highest-order modes is because they have fewer guided nearest-neighbours (and so are coupled more strongly to cladding modes). The same calculations are shown for the typical 62.5 μm GRIN fibre in Supplementary Figures 3-6.

**Theoretical Coupling Model**

In general, the coupling between two of the fibre's eigenmodes, $\varphi_l$ and $\varphi_m$, caused by a general local index perturbation, $\Delta n(x, y, z)$, is of the form[2]:

$$C_{lm}(z) \propto \iint dxdy \Delta n(x, y, z) n_o(x, y) \varphi_l^*(x, y) \varphi_m(x, y)$$

$$= \iint dxdy \Delta n(x, y) n_o(x, y) \varphi_l^*(x, y) \varphi_m(x, y) \cdot f(z) \qquad (1)$$

where $n_o(x, y)$ is the ideal fiber's refractive index profile. If we factor $\Delta n(x, y, z)$ into its transverse and longitudinal components, $\Delta n(x, y, z) = \Delta n(x, y) \cdot f(z)$, and we neglect the influence of all other modes, the amplitude of mode $l$ at a given position, is proportional to $F(K = \Delta\beta)$. $F(K)$ is the Fourier transform of $f(z)$, evaluated at the propagation constant mismatch, $\Delta\beta = \beta_m - \beta_l$, between the two modes. In general, the energy exchanged between $\varphi_l$ and $\varphi_m$ will be larger when $C_{lm}(z)$ is larger. Hence, the mode coupling depends on the spatial overlap between the modes and the perturbation, as well as the longitudinal evolution of the perturbation, $f(z)$. With two modes and $f(z) = 1$, energy is periodically exchanged. With many modes and/or with more complex $f(z)$, energy transferred from $\varphi_m$ to $\varphi_l$ may never return to $\varphi_m$ (it may stay in $\varphi_l$ or be coupled to another mode, $\varphi_k$).

*Nonlinearity*

For the Kerr nonlinearity, $\Delta n(x, y, z) = n_2 I(x, y, z)$, where $I(x, y, z) = \left|\sum_{m=1}^{N} a_m(z) \varphi_m(x, y) e^{-i\beta_m z}\right|^2$ is the total field intensity. The field itself therefore determines the coupling (Eqn. 1), which recursively affects its own propagation and subsequent coupling. We can nonetheless use the mode-coupling network to understand the nonlinear dynamics. In a waveguide with a parabolic index profile, the average intensity, $\frac{1}{L}\int_{z=0}^{L} dz\, I(x, y, z)$, will be bell-shaped, and peaked at the centre of the fibre. This is because light tends to be more concentrated in regions with higher refractive index. Since low-order modes (LOMs) are more localized, they have a higher nonlinear overlap integral with all other modes. Hence on average, the coefficients at a given position, $z'$, in the fibre, $C_{lm}(z')$, will be biased to LOMs (Fig. 2a-b, Supplementary Figures 3-4).

The longitudinal evolution, $f(z)$, of the index perturbation caused by the Kerr nonlinearity includes periodic and aperiodic components. Modes in a parabolic-index fibre are clustered into groups (Fig. 2c, Supplementary Figure 5) with similar propagation constants, and the number of modes in each group increases with decreasing propagation constant. The separation between these groups is constant, so a multimode field undergoes a linear, *periodic* self-imaging. Since the intensity $I(x, y, z)$ depends recursively on itself, *aperiodic* evolution along $z$ also occurs. Therefore, $F(K)$ is peaked near $K = 0$ and at harmonics of the self-imaging frequency, so nonlinear coupling within and between groups is expected.

High-order modes (HOMs) in a GRIN fibre are susceptible to instabilities that break their symmetries[35]. This follows from the mode-coupling structure: HOMs have complex symmetries and many neighbouring modes with $\Delta\beta$ near 0. As a result, HOM nonlinear dynamics are complicated. However, on average, energy is more likely to couple from HOMs to LOMs than vice versa due to the bias of $C_{lm}(z)$. Since this causes $I(x, y, z)$ to overlap even more with LOMs, the bias and energy in LOMs both grow monotonically. Meanwhile, as the sole member of its group, the fundamental mode is nonlinearly stable[35], so energy can accumulate within it.

*Disorder*

Fibres contain small index perturbations that primarily couple modes with similar propagation constants. This disorder arises from stochastic density fluctuations of the glass and dopants, as well as small manufacturing errors and environmental effects such as bending, twisting, and core ellipticity. These small irregularities lead to scattering, which increases rapidly with decreasing wavelength. Hence while the net effect of disorder is still small, it is much stronger for our 532-nm experiments than for longer wavelengths. These random perturbations are longitudinally aperiodic, so $F(K)$ is centred around $K = 0$. As a result, disorder causes coupling between modes with similar propagation constants; primarily within mode groups[2,5,6]. Since most sources of disorder are asymmetric, $C_{lm}(z)$ is usually large when $\Delta n(x, y)$ corresponds to a symmetry between the two modes. Since the perturbations are small, $C_{lm}(z)$ is also large only for similarly-sized modes (Supplementary Figure 2, 5-6). As a consequence of these characteristics, the fundamental mode is least affected by disorder, while HOMs, which have reduced symmetries and many neighbours, are most affected (Fig. 2c, Supplementary Figures 2,6).

*Dissipation*

Linear loss and nonlinear loss occur during propagation. Linear loss due to coupling to cladding modes most strongly affects HOMs, since their propagation constants are closest to those of cladding modes. Stimulated Raman scattering, a nonlinear dissipative process, transfers energy to redshifted Stokes waves in proportion to the overlap between the pump and the Stokes field. If the pump occupies a single pure mode, the Stokes field will be dominated by the same mode. However, for a multimode pump, the Stokes modes will differ from the pump modes due to gain competition. Since the average intensity of a multimode field in a parabolic fibre overlaps most with the fundamental mode, the fundamental mode dominates for most pump fields [32-33].

*Total Coupling Effects*

In our experiments, mode coupling is the combination of the couplings caused by nonlinearity, disorder, and dissipation. Raman beam clean-up is only the most probable outcome of Raman gain competition in multimode GRIN fiber[32]. Meanwhile, disorder alone is diffusive: it causes additional dissipation of HOMs, but otherwise leads to energy equipartition. In the absence of other effects, the Kerr nonlinearity alone can lead to the observed attractor for some initial conditions[16-19]. However, our experiments include substantial disorder and dissipation. Moreover, the attractor is observed for virtually all coupling conditions possible with a near-Gaussian beam, provided sufficient energy (typically 2-3 µJ) is coupled into the fibre.

The interaction between coupling mechanisms is important. Through its influence on dissipation and nonlinear coupling, weak disorder enhances the attractor (Supplementary Figures 7-14). Disorder couples HOMs to cladding modes, thereby increasing their attenuation. This increases the energy in LOMs relative to HOMs. As a result, the overlap of the nonlinear coupling integral with those modes is increased, and the bias of $C_{lm}(z)$ is enhanced.

Meanwhile, disorder-induced coupling within mode groups also interacts with nonlinearity. If disorder overwhelms nonlinearity, it can completely suppress intermodal (or intergroup) energy exchange[26-27,29]. However, for the weaker coupling here, random coupling enhances the attractor. By averaging over HOMs, Raman beam cleanup is more likely to occur (as was suggested by Chiang[33-34]). By seeding perturbations, disorder lowers the power threshold for nonlinear HOM instability (Supplementary Figures 11-14). Since the fundamental mode is spatially stable and least affected by disorder, this is an important process underlying the attractor for the ~ 1 kW peak powers considered here.

*Spatiotemporal Instability*

The fundamental origin of the observed spectral sidebands is spatiotemporal modulation instability (STMI). STMI consists of multiple four-wave mixing processes with spatiotemporal phase-matching: spatial (modal) dispersion compensates for chromatic dispersion, as in STMI in free space[49]. Fig. 3 shows theoretical calculations along with experimental measurements of STMI. Since STMI sidebands overlap with the red-most edge of the Raman cascade, the Raman peaks blur together. Furthermore, since there are several intense Stokes waves, STMI sidebands may originate from several different pump wavelengths. With the attractor as the pump, the higher-order sidebands correspond to increasingly higher-order mode groups (Fig. 3a). Fig. 3b shows the measured spectrum, and Fig. 3c the spatial profiles measured at the indicated frequencies, which exhibit the theoretical spatiospectral trend. Figure 3d shows the typical locations of the MI sidebands (for the fibre used in Fig. 1) along with theory.

Despite its 2D stability, the attractor is the maximally unstable state of the field when its full 3D nature is considered. Fig. 4, Supplementary Movies 2-3, and Supplementary Figures 15-20 demonstrate the maximal instability of the attractor through numerical simulations. The fact that the nonlinear attractor here is the most unstable, or critical state, is related to the broader concept of self-organized criticality[46-47].

Understanding the instability of the attractor requires consideration of time-domain processes. The attractor is driven by spatial processes; it is primarily a 2D, narrowband process. However, the field is 3D, and for sufficiently intense and/or broadband fields, spatiotemporal instabilities become important. Intense fields may become susceptible to spatiotemporal instabilities before reaching the attractor[12-13], but they are particularly relevant once the attractor is reached, since the attractor is the maximally unstable state. Rigorous understanding of the coupling between different frequencies and modes will require a generalization of the monochromatic approach used so far[48]. Nonetheless, we may simply extend the existing approach to include chromatic dispersion, letting $\Delta\beta = \beta_m(\omega_m) - \beta_l(\omega_l)$, in order to obtain an intuitive understanding of observed phenomena. The qualities that make the attractor two-dimensionally stable make it three-dimensionally unstable: it has the highest intensity, and being the sole member of its group, is unaffected by the intergroup coupling suppression that arises from nonlinear and disorder-induced intragroup coupling. In our experiments, the instability develops from the attractor, for which periodic oscillations are small. Therefore $F(K)$ is large only near $K = 0$, and coupling between different modes can only take place if they have different frequencies. This differs from geometric parametric instability (GPI)[16,18]. In terms of our qualitative model for spatiotemporal coupling, GPI results from strong periodic oscillations, which create high-amplitude harmonics in $F(K)$, allowing coupling between the same modes at different frequencies. We therefore primarily observe STMI, which is evident from the hyperbolic spatiospectral profile shown in Fig. 3.

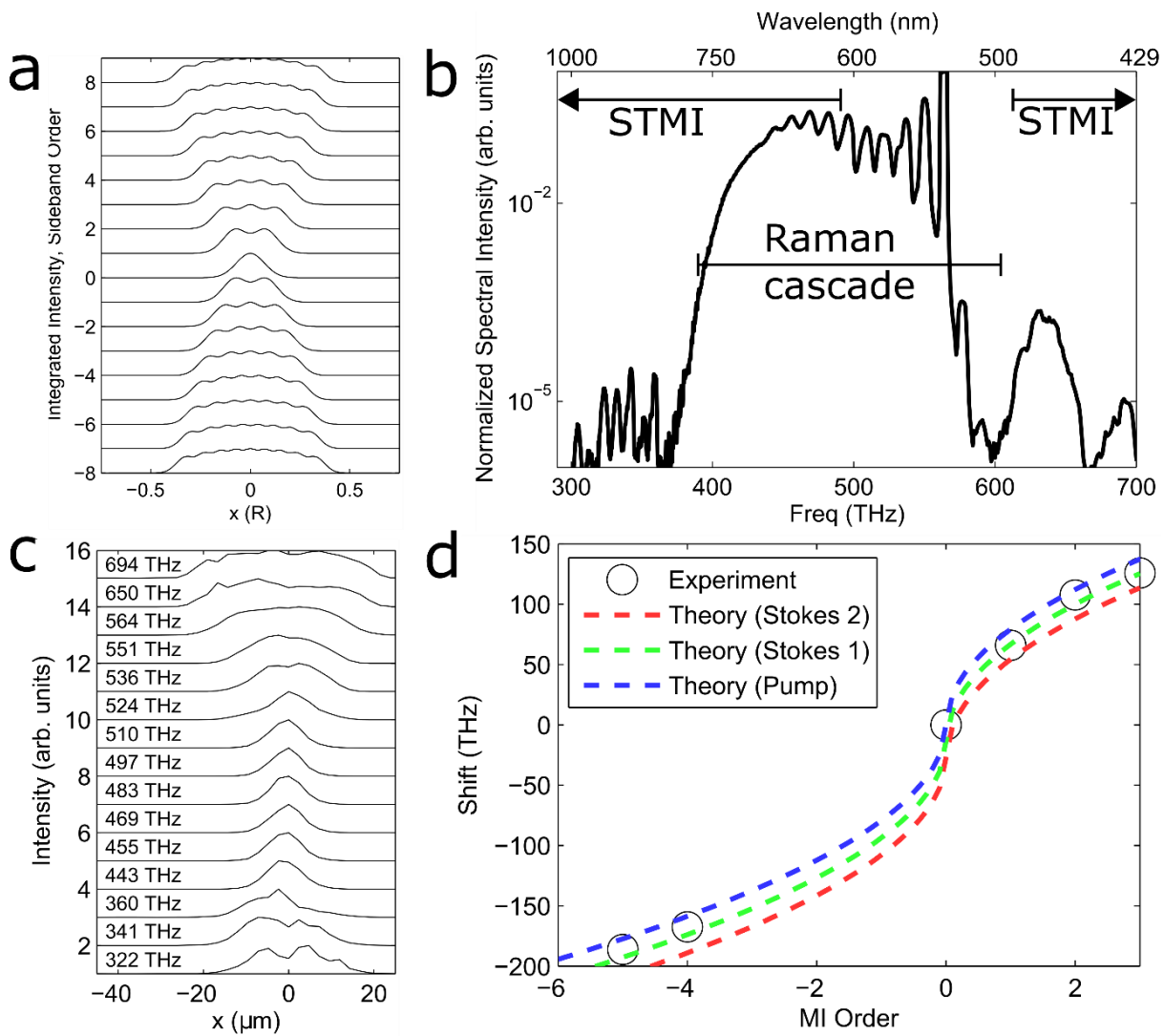

**Figure 3: Spatiotemporal modulation instability.** (a) Theoretical spatial profile, integrated over one spatial dimension, for increasing orders of modulation instability (R is the fibre core

radius). (b) Spectrum of the entire field, with the salient features indicated. STMI: spatiotemporal modulation instability sidebands. (c) Experimentally-measured spatial profiles at the indicated frequencies. Note that, because cascaded Raman also leads to spatiospectral structure, and because multiple pumps lead to MI sidebands, the comparison is not direct. Rather, it is meant to illustrate the overall similar hyperbolic shape. (d) Average position of measured MI sidebands (for 5 different initial conditions) compared to theory from Ref. 5. The theoretical curves are shown for the 532-nm pump and two Stokes waves (545 nm, 559 nm).

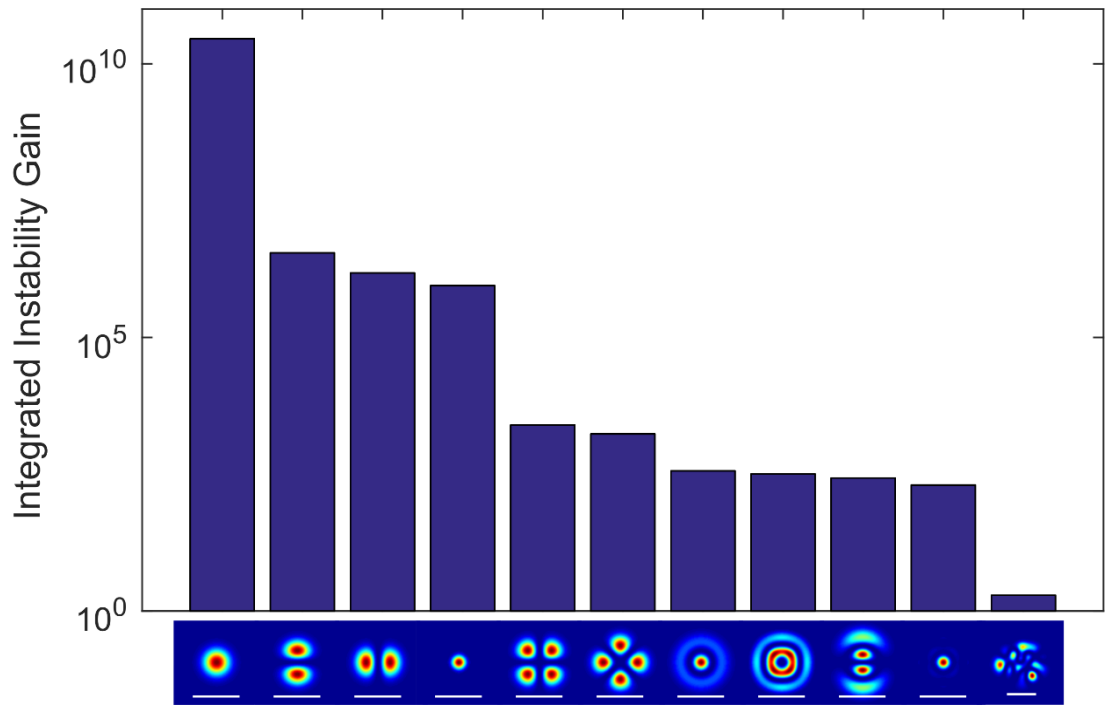

**Figure 4: Maximal instability of the attractor.** We conducted numerical simulations with the first 30 modes of the fibre. In order to excite spatiotemporal instabilities rapidly, we used a peak power of 560 kW, which is still far below the critical power for self-focusing at 532 nm. The y-axis (log scale) shows the integrated instability gain for the first redshifted sideband (the energy in the sideband relative to the initial noise energy) for the initial spatial conditions shown below. The attractor produces the most instability gain by a wide margin (~$10^4$ more than the second initial condition). This conclusion is supported by simulations with lower peak power (Supplementary Figures 15-20). The scale bars for the lower plots are 11 µm.

## Discussion

While the unstable attractor we study here is of particular interest due to its multifaceted origin, its basic features appear to be quite universal, manifesting across a broad range of parameters. Here, we observe self-focusing towards a maximally-unstable attractor, followed by rapid development of spatiotemporal modulation instability. This occurs with kilowatt peak power, nanosecond-duration pulses in 100-m, normal dispersion fibres with 55 modes. Qualitatively similar behaviour has been observed in shorter fibres with many more modes (~100-300), with higher power and normal-dispersion[16-18], and with up to megawatt peak power, 100-fs duration pulses in short anomalous-dispersion fibres[12,13]. This universality stems from the common characteristics of the mode coupling network.

As we have outlined, mode-coupling in GRIN fibre is primarily local, between modes with similar propagation constants (intragroup coupling), with some intergroup coupling. This is a small-world network[36]: a network where nodes are primarily connected locally, to their near-neighbours, with some small number of 'shortcut' connections between the neighbourhoods. Small-world networks exhibit a phase transition to global interconnectivity when the strength of shortcut coupling exceeds a threshold. Considering disorder only, intergroup coupling is crucial for observing a phase transition from dispersive to diffusive propagation[5], and for suppression of four-wave mixing[29]. With nonlinearity, intergroup coupling particularly arises from the periodic component of $f(z)$, and from broadband FWM processes like STMI. In the attractor, this suggests a crucial role for the lowest-order modes in each mode group. These modes, as can be seen in Fig. 2a, are strongly coupled to one another, making them intergroup coupling 'hubs'. Shortcut coupling transitions may underlie the 2D attractor, the growth of spatiotemporal complexity following STMI, and the emergence of ultrabroadband spatiotemporal coherence through the geometric parametric instability[16-18].

The confluence of different intermodal coupling processes and emergent behaviour of many modes observed here is a compelling example that recommends MM waveguides as an experimental test bed for complexity science. This use could be similar to single-mode fiber[50], but MMFs have many more degrees of freedom and control, and therefore a capacity to connect to a much broader class of complex dynamical systems. Furthermore, this capacity also makes MM waveguides platforms to realize particular applications of those fields, such as the neural network computers that have experienced a resurgence of attention.

For other applications, multimode fibres have significant potential but it is still not clear if nonlinear processes can be controlled in useful, reliable ways. Beam clean-up appears to provide one route to spatially-coherent multimode sources, but the instability of the attractor may undermine this potential. The worldwide demand for increased telecommunications bandwidth at decreasing cost poses an obstacle that SMF-based systems are fundamentally unable to avoid. Indications are that by 2020, SMF transmission capacity will be two orders of magnitude less than anticipated needs[3]. The performance limits of MMF-based transmission are still the subject of research, and complex intermodal interactions are a potential obstacle. Here we have observed collective dynamics due to the cooperation of disorder and dissipation with nonlinearity. We use relatively low peak-power (~1 kW) and our experiments roughly compare to a loss-managed line of 150 km (Supplementary Table 1). This highlights the potential complexity of MMF transmission impairment, and the importance of understanding not only nonlinear interference effects between modes, but also the possibility of collective dynamics involving many channels. Ultimately, realistic multimode telecommunications scenarios will involve disorder, dissipation, and nonlinearity. Understanding their individual and collective effects is an important area for future research.

We have shown how a given initial field in graded-index multimode fibre self-organizes into a state which is most unstable through the cooperation of nonlinearity, dissipation and disorder. The subsequent evolution from this state causes the field to develop a spatiotemporally-complex nature characterized by a high degree of space-time coupling. The results showcase the wide-ranging complex phenomena that can be investigated effectively in MMF and raise many questions and opportunities. Applications include powerful, flexible fibre lasers, space-division multiplexing, and novel computing platforms.

## Methods

### Main experiments

We conducted experiments by launching ~1-ns pulses from a Q-switched, frequency-doubled Nd:YAG laser into GRIN fibres with lengths 50-100 m. The fibres have a numerical aperture of 0.137, a radius of 13 µm and are nearly parabolic ($\alpha = 1.95$). The small core translates to enhanced nonlinearity, and to relatively few modes, thereby increasing the validity of the simulations. The initial spatial condition (i.e., combination of modes excited) was varied by adjusting the position of the fibre using a precision 3-axis stage, and the launched power was varied using a waveplate and polarizer. For the majority of initial conditions, we observed results similar to those shown in Fig. 1 and Fig. 3. In exceptional cases, STMI may develop before the field reaches the ultimate state of the attractor (see for example Supplementary Figure 23-24). However, in these cases self-organized instability and evolution towards the attractor are still observed.

### Mode coupling matrices

For Fig. 2a-b, showing the nonlinear mode-coupling matrix, we computed the coefficients from Eqn. (1), at a particular $z$-value (i.e., $C_{lm}(z')$) for 500 random fields. These fields are each generated as the coherent sum of random combinations of the fibre's 55 modes, i.e., $I(x,y) = \left|\sum_{l=1}^{55} c_l \varphi_l\right|^2$, where $c_l = a_l + ib_l$ and $a_l$ and $b_l$ are random numbers from a uniform distribution from -1 to 1. This is therefore the average mode-coupling matrix for a random field in the fibre at a given position $z'$ along the fibre. This illustrates the typical characteristics of the nonlinear mode-coupling.

For the disorder-induced mode-coupling matrix (Fig. 2c, Supplementary Figure 1-2), we computed the coefficients from Eqn. (1) for 100 random bends of the fibre, described by $\Delta n(x,y) = \frac{\Delta_x x}{R} + \frac{\Delta_y y}{R}$ for $\sqrt{x^2 + y^2} \leq R$ and = 0 otherwise. The values $\Delta_{x,y}$ were obtained from a uniform distribution ranging from $-0.025 * \Delta$ to $0.025 * \Delta$, where $\Delta$ is the difference between the center and cladding index of the fibre, 0.0064.

### Spatiospectral measurement

We inserted a cylindrical lens at the end of the 4-f telescope used to image the near-field beam profile, so that the focus of the cylindrical lens coincided with the imaging plane of the telescope. A bare fibre was scanned through the beam along the dimension unfocused by the cylindrical lens. In order to observe the clearest hyperbolic spatiospectral shape, it is helpful to have a large size difference between the modes. Therefore we chose a fibre which has a larger core size (≈ 20 μm radius), but similar numerical aperture, as the one used for the experiment shown in Fig. 1. Similar measurements as shown in Fig. 3 for the fibre used in Fig. 1 are shown in Supplementary Figures 21 and 22. Furthermore, for comparison to theory in Fig. 3a, we used parameters of typical 62.5 μm core diameter GRIN fibre, which has more modes than the fibres studied, so that the plot could be extended to high mode groups to show the trend. The modes are plotted integrated over one spatial dimension in order to facilitate comparison with experiment.

**Instability simulations**

We conducted simulations using the generalized multimode nonlinear Schrödinger equation[11]. We considered the first 30 linearly-polarized modes of the fibre, which were numerically calculated. In order to ensure that spatiotemporal instabilities were excited and that the attractor was not important to the evolution, and in order to conduct the simulations in a timely fashion, we launched high-power (560 kW) pulses. This is $0.45 * P_{crit}$ at 532 nm. The launched pulses were 3.2 uJ, 5 ps Gaussians, with the initial spatial profiles shown in Figure 4. The sideband energies were measured after 0.6 cm, which was sufficient to see substantial gain for most initial conditions. In order to make sure the results were not the result of the large peak power, several selected initial conditions were tested at lower power ($0.03 * P_{crit}$) and longer propagation lengths (>30 cm). These simulations yield the same conclusion as the high-power ones (Supplementary Figures 15-20).

**Acknowledgments**

Portions of this work were funded by Office of Naval Research grant N00014-13-1-0649 and by National Science Foundation grant ECCS-1609129. We thank OFS for providing some of the fibre used in the experiments. We thank Z. Zhu, K. Krupa, A. Tonello, A. Barthélémy, V. Couderc, G. Millot, and S. Wabnitz for discussions.

**Author Affiliations**

L.G.W, Z.L. and F.W.W.

School of Applied and Engineering Physics, Cornell University, Ithaca, New York 14853, USA

D.A.N and M.-J. L

Corning Incorporated, Sullivan Park, Corning, NY 14830, USA

D.N.C

CREOL, College of Optics and Photonics, University of Central Florida, Orlando, Florida 32816, USA

**Author Contributions**

L.G.W. performed simulations and experiments, with assistance provided by Z.L. D.A.N and M.-J. L. made and provided small-core GRIN fibres. L.G.W. and F.W.W wrote the first drafts of the manuscript and all authors contributed to the final version.

**Competing financial interests**

D.A.N and M.-J. L are employed by Corning Incorporated, which manufactures optical fibre for applications including telecommunications.

**Materials & Correspondence**

Correspondence and requests for materials to: L.G. Wright lgw32@cornell.edu

# Supplementary Information for "Self-organized instability in graded-index multimode fibres"

Logan G. Wright[1], Zhanwei Liu[1], Daniel A. Nolan[2], Ming-Jun Li[2], Demetrios N. Christodoulides[3], and Frank W. Wise[1]

1. School of Applied and Engineering Physics, Cornell University, Ithaca, New York 14853, USA
2. Corning Incorporated, Sullivan Park, Corning, NY 14830, USA
3. CREOL, College of Optics and Photonics, University of Central Florida, Orlando, Florida 32816, USA

Contents

Supplementary Figures 1-6: Quantification of nonlinear and disorder-induced coupling in GRIN MMFs

Supplementary Figures 7-14: Numerical verification of the role of disorder in beam clean-up effects

Supplementary Figures 15-20: Numerical validation of the attractor's maximal instability

Supplementary Figures 21-22: Examples of more advanced supercontinuum spatiospectral measurements

Supplementary Figures 23-24: Experimental results for an exceptional case of evolution where the field does not completely reach the attractor

Supplementary Table 1: Comparison of nonlinearity in experiment to typical fibre transmission line

Movies

Supplementary Movie 1: Animated version of Figure 1

Supplementary Movie 2: Mode-resolved simulation of propagation of the attractor, with high peak power, exciting MI sidebands.

Supplementary Movie 3: Same as Supplementary Movie 2, except plotted in the space-time domain, integrated over one spatial dimension.

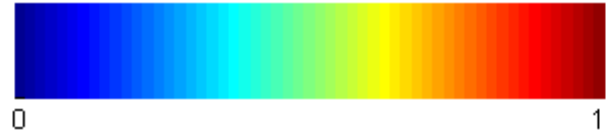

Supplementary Figure 1: Colour scale for mode-coupling matrices. Values are plotted in terms of their magnitude relative to the maximum magnitude, $|C_{nm}(z')|/|C_{nm}(z')|_{max}$.

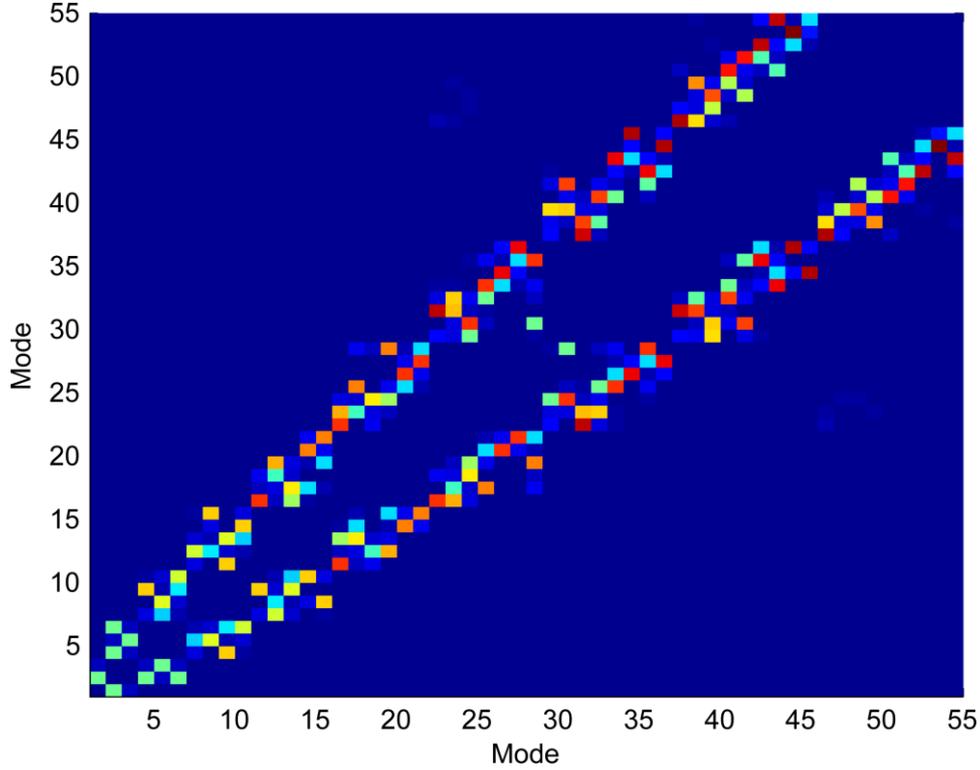

Supplementary Figure 2: Example disorder-induced mode-coupling matrix for the 26 µm diameter fibre used in the experiment. We computed the coefficients from Eqn. (1) for 100 random bends of the fibre, described by $\Delta n(x,y) = \frac{\Delta_x x}{R} + \frac{\Delta_y y}{R}$ for $\sqrt{x^2 + y^2} \leq R$ and $= 0$ otherwise. The values $\Delta_{x,y}$ were obtained from a uniform distribution ranging from $-0.025 * \Delta$ to $0.025 * \Delta$, where $\Delta$ is the difference between the centre and cladding index of the fibre, 0.0064.

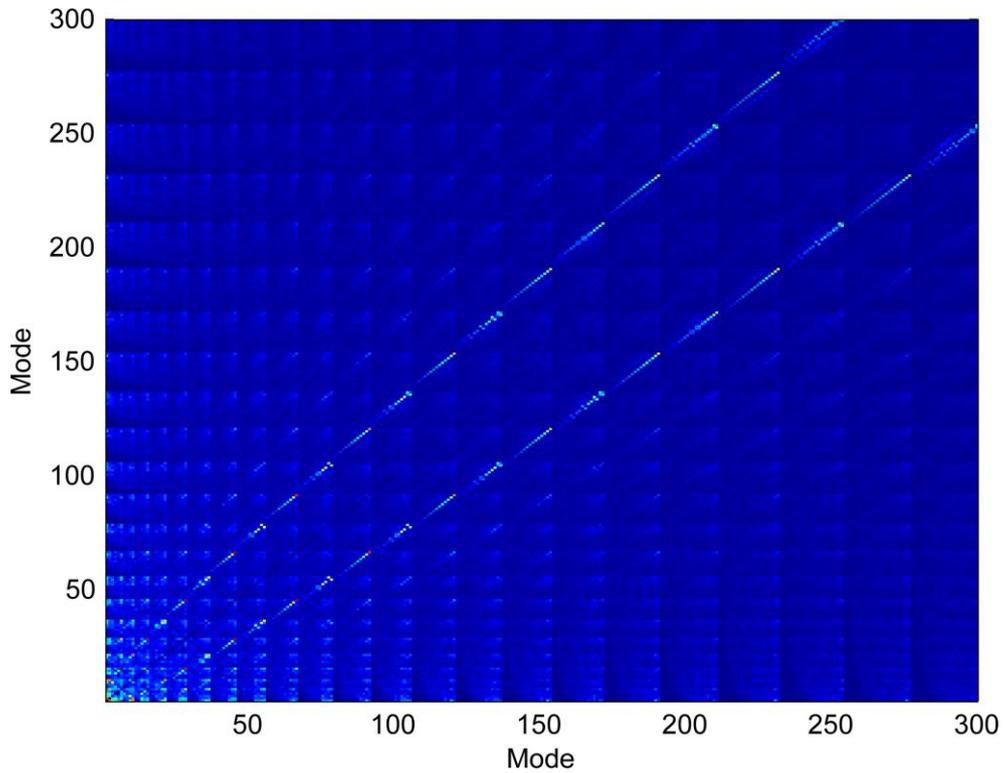

Supplementary Figure 3: The mode-coupling matrix at particular $z$-value, $C_{nm}(z')$, due to Kerr nonlinearity, averaged for 12 random combinations of the first 300 modes of a commercial 62.5 um diameter, 0.275 NA GRIN fibre at 1030 nm

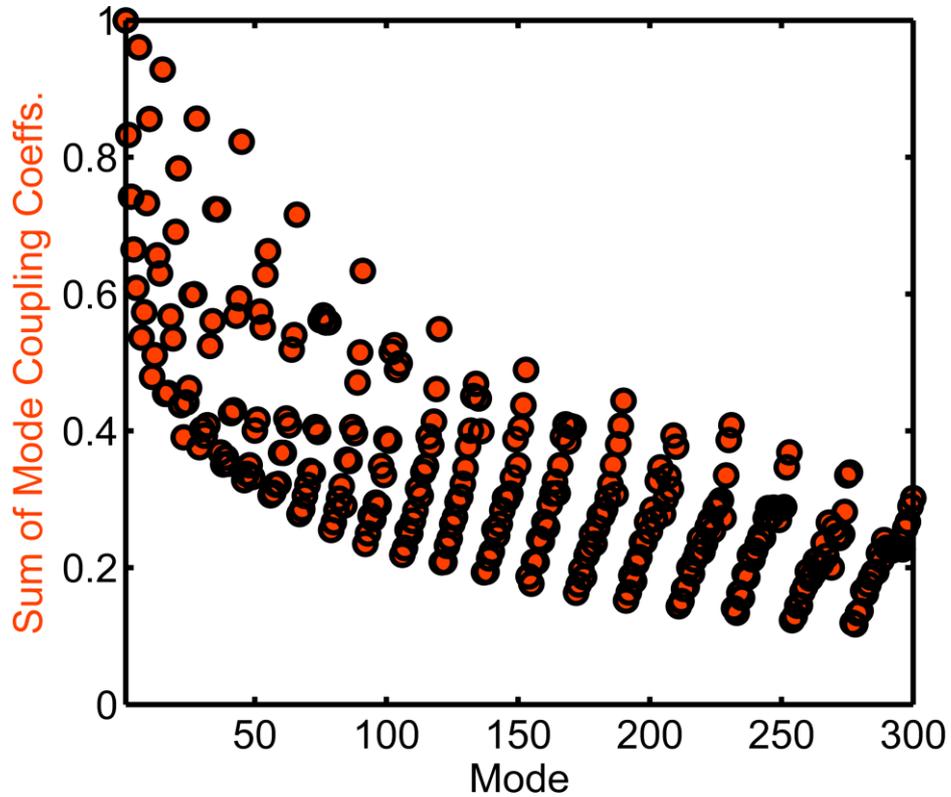

Supplementary Figure 4: The sum of the mode-coupling coefficients for each mode in 62.5 μm diameter, 0.275 NA GRIN fibre. Calculated for the first 300 modes at 1030 nm

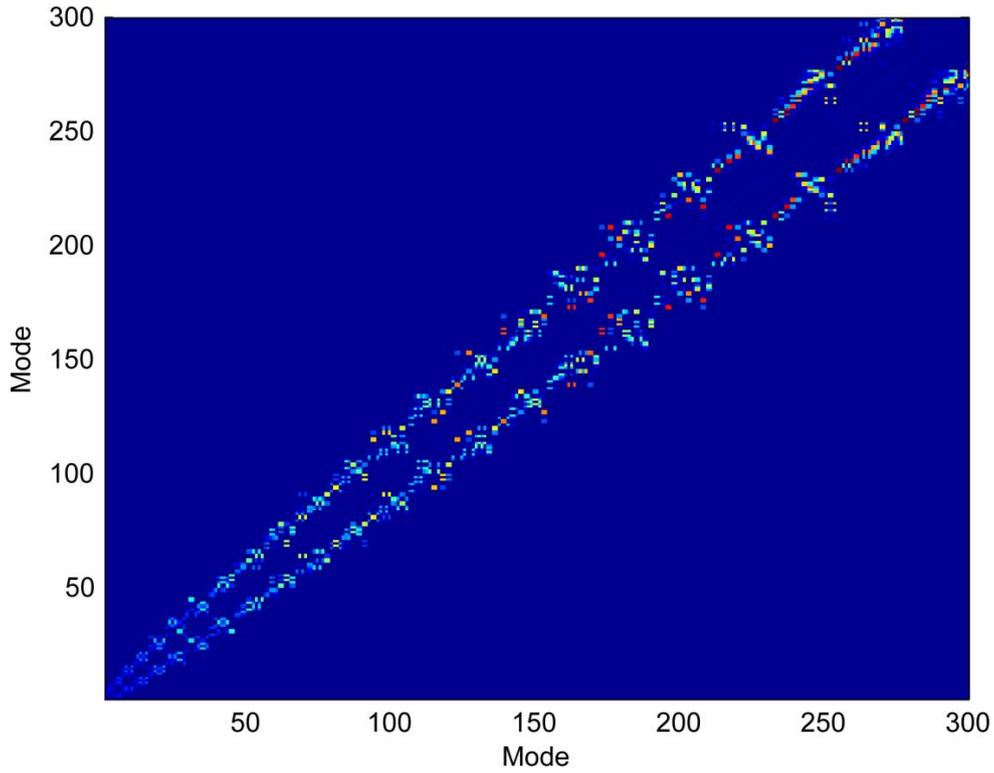

Supplementary Figure 5: Example disorder-induced mode-coupling matrix for GIF625 at 1030 nm used in the experiment. We computed the coefficients from Eqn. (1) for 10 random bends of the fibre, described by $\Delta n(x,y) = \frac{\Delta_x x}{R} + \frac{\Delta_y y}{R}$ for $\sqrt{x^2 + y^2} \leq R$ and $= 0$ otherwise. The values $\Delta_{x,y}$ were obtained from a uniform distribution ranging from $-0.025 * \Delta$ to $0.025 * \Delta$, where $\Delta$ is the difference between the centre and cladding index of the fibre, 0.0264.

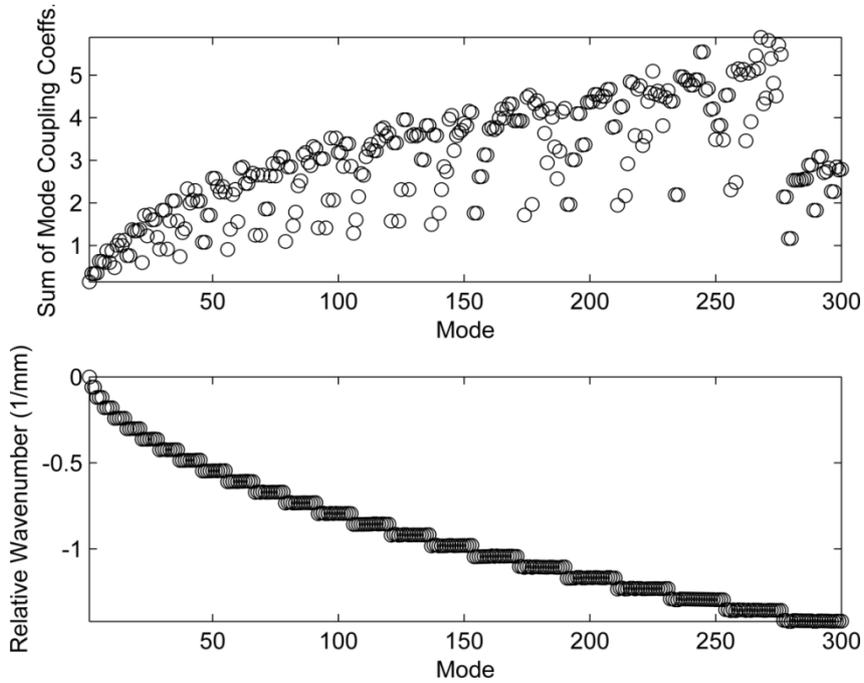

Supplementary Figure 6: (top) the sum of the mode-coupling coefficients for each mode, and (bottom) the wavenumber (propagation constant) for each mode, relative to the fundamental mode, in 62.5 μm diameter, 0.275 NA GRIN fibre. Calculated for the first 300 modes at 1030 nm

We launched combinations of the 55 modes of the fibre used for experiments shown in Figure 1. To represent disorder, we use small bends. While far from a complete description of disorder in fibre, the implementation here captures the main features of disorder as pertains to mode-coupling in multimode fibre: it constitutes a small, symmetry-breaking perturbation. Bending introduces a shift to the index profile that varies linearly across the fibre. Each bend therefore results in an index perturbation that can be approximated as $\Delta n(x,y) = \frac{\Delta_x x}{R} + \frac{\Delta_y y}{R}$. The values $\Delta_{x,y}$ were obtained from a uniform distribution ranging from -0.025·0.0064 to 0.025·0.0064, where 0.0064 is the different between the core and cladding index. To ensure that the disorder was aperiodic, the values $\Delta_x$ and $\Delta_y$ were updated every $x_1$ and $x_2$ cm, where $x_1$ and $x_2$ are random numbers from a uniform distribution *[2,6]* as well as every 10 cm. Note that this is very long compared to intergroup beat lengths in the fibre (around 0.5 mm). The longitudinal resolution used was 1 μm, and the transverse grid was 256x256 with a range of 49.2 μm.

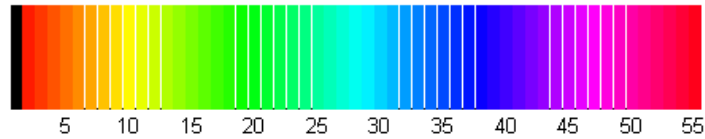

Supplementary Figure 7: Colour coding of modes in Supplementary Figures 8, 11-14

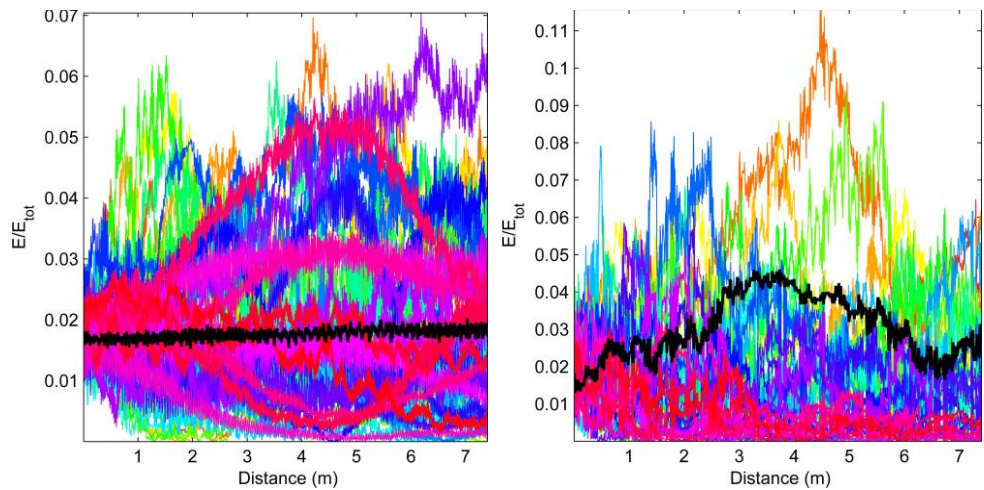

Supplementary Figure 8: The evolution of energy in the modes for a simulation of a 10-kW field launched equally into the first 55 modes of the fibre without disorder (left) and with disorder (right). With only nonlinearity, energy builds up only very slowly into the fundamental mode (black). With disorder, energy increases much more rapidly. For both disordered and disorder-free fibres, the growth of the fundamental mode is not monotonic. In particular, the fundamental exchanges energy with other dominant low-order modes, so its energy increases on average. However, disorder increases the flow of energy from high- to low-order modes (see below).

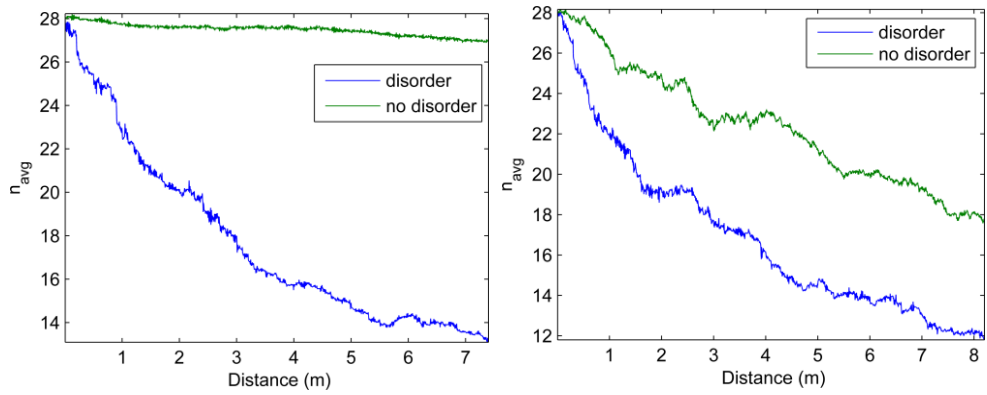

Supplementary Figure 9: The energy-weighted average mode number for simulations of a 10-kW field (left) and 50-kW field (right) launched equally into all 55 modes. Disorder decreases the average mode number. With increased nonlinearity, the importance of disorder is less important.

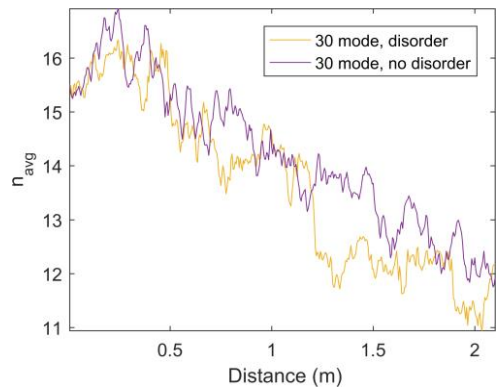

Supplementary Figure 10: The energy-weighted average mode number for simulations of 150-kW fields launched equally into the first 30 modes. The effect of disorder is evidently smaller for relatively lower-order modes and for higher nonlinearity.

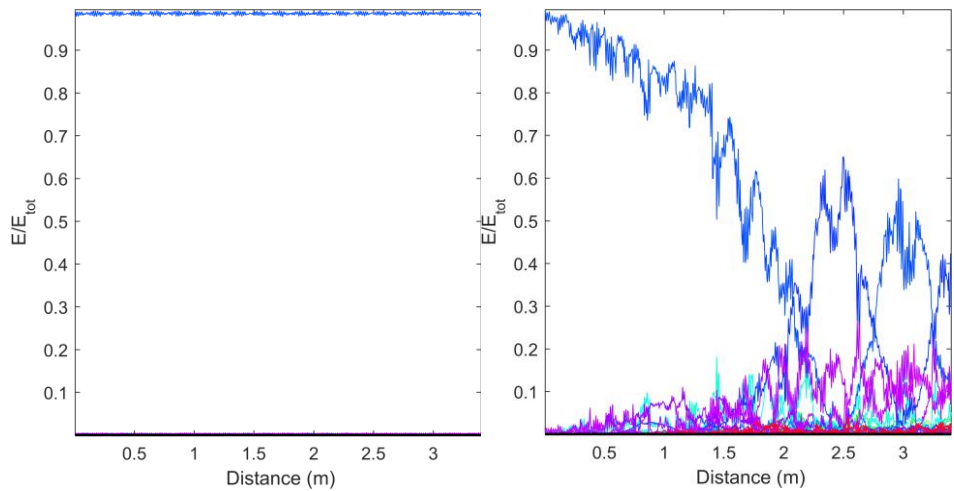

Supplementary Figure 11: Propagation of a 50-kW field launched into the 35th mode of the fibre. Left: propagation without disorder. Right: propagation with disorder. Disorder is important for lowering the instability threshold power for HOMs. It is therefore crucial for HOM instability in the low-power regime of our experiments. HOM instability and disorder lead to rapid energy exchange within mode groups and nearby modes. This increased local coupling can interfere with coupling between groups, since energy tends to couple locally before it can couple elsewhere. This interference affects the more-populous HOM groups more than LOM groups. As a result, even though it decreases the likelihood of intergroup coupling, it increases the likelihood that energy will *permanently* transfer from high- to low-order modes.

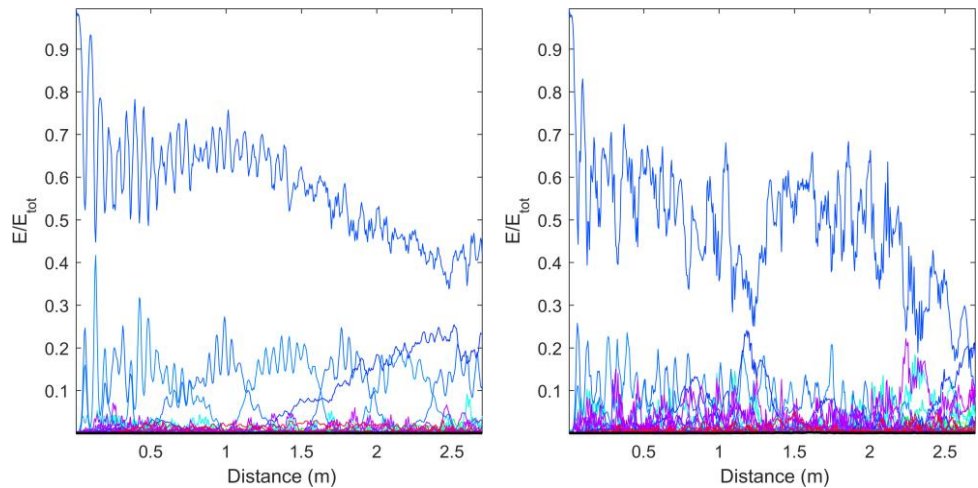

Supplementary Figure 12: Propagation of a 150-kW field launched into the 35$^{th}$ mode of the fibre. Left: propagation without disorder. Right: propagation with disorder. As the power is increased, the relative importance of disorder is lessened, and the propagation is similar.

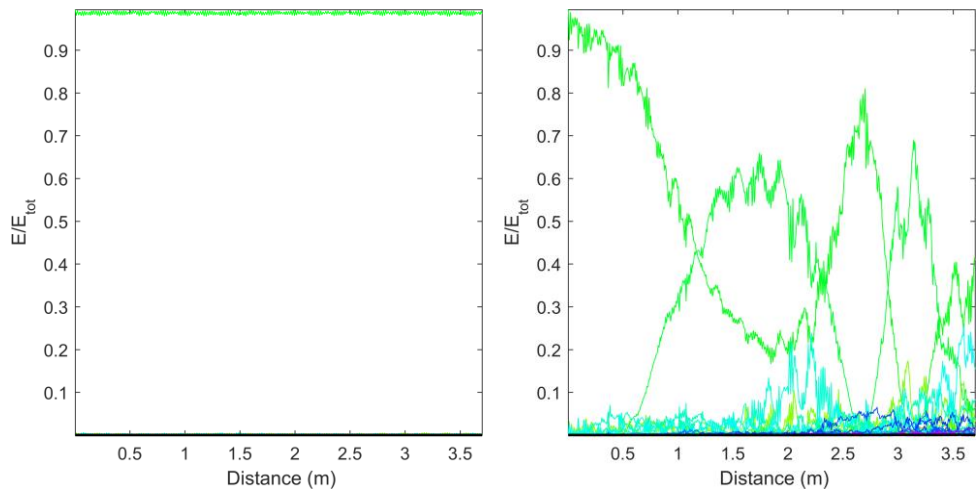

Supplementary Figure 13: Propagation of a 50-kW field launched into the 20$^{th}$ mode of the fibre. Left: propagation without disorder. Right: propagation with disorder. This provides another example of disorder-enhanced HOM instability.

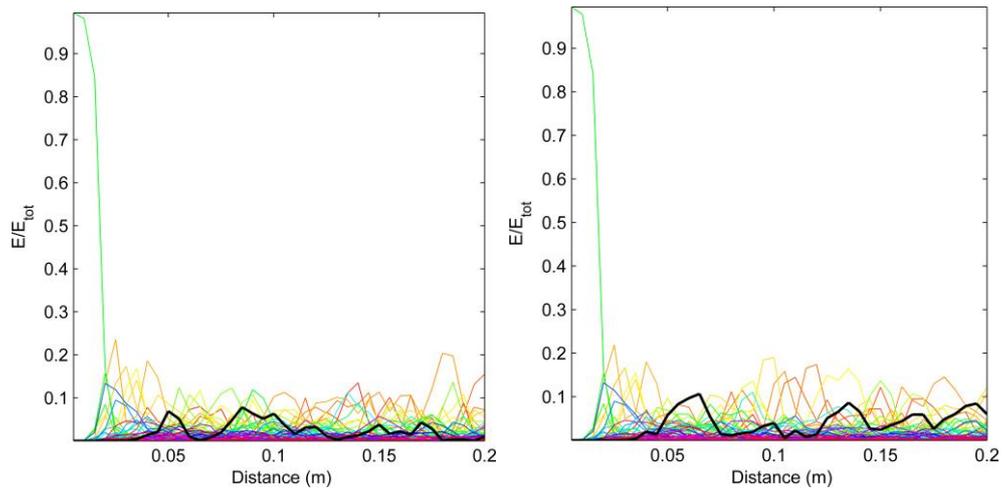

Supplementary Figure 14: Propagation of a 500-kW field launched into the 20$^{th}$ mode of the fibre. Left: propagation without disorder. Right: propagation with disorder. With higher power, the role of disorder on the instability of the HOM is virtually negligible.

To evaluate the maximal instability of the attractor, we simulated the propagation of different initial conditions in a regime with higher peak power than in the experiment. Due to the large peak power, spatiotemporal modulation instability can be excited with many different initial conditions. To evaluate the stability of each initial condition, we observe the rate at which MI sidebands develop for a fixed energy. Simulations were performed using the generalized multimode nonlinear Schrödinger equation with the first 10 and with the first 30 modes of the fibre used in the experiment. For both sets of modes, we observe that the attractor exhibits substantially faster growth of sidebands than any other initial condition.

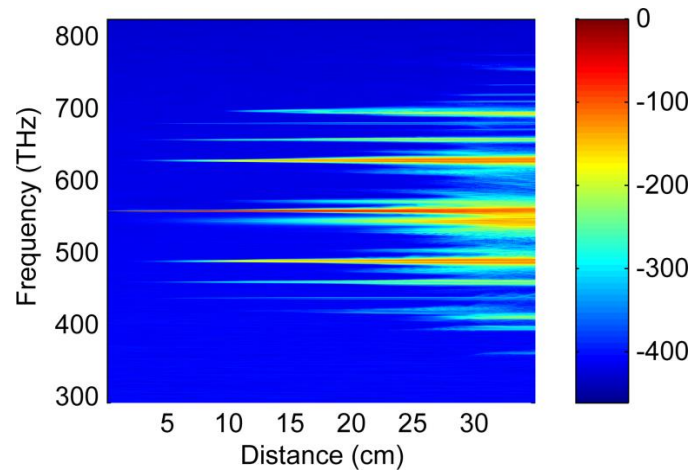

Supplementary Figure 15: Spectral evolution of the field launched into the fibre used in the experiment, for a 7-ps, 320-nJ pulse. The simulation is conducted using the generalized multimode nonlinear Schrödinger equation, with the first 10 modes of the fibre at 532 nm. The field is initially in the fundamental mode, with a background of the other modes, each at 0.01% the energy launched into the fundamental mode. The rate of growth of the MI sidebands is much greater than for any of the other initial conditions tested, confirming the expectation that the fundamental mode (with a background of higher-order modes) is the most unstable state.

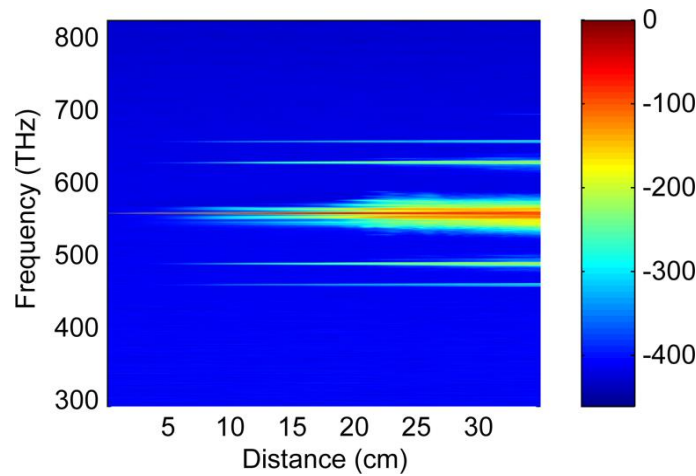

Supplementary Figure 16: Spectral evolution of the field launched into the fibre used in the experiment, for a 7-ps, 320-nJ pulse. The simulation is conducted using the generalized multimode nonlinear Schrödinger equation, with the first 10 modes of the fibre at 532 nm. The field is initially in the 2nd mode, with a background of the other modes, each at 0.01% the energy launched into the 2nd mode.

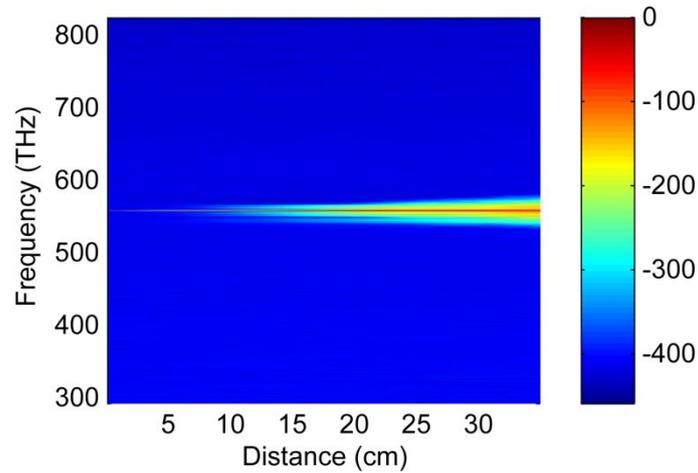

Supplementary Figure 17: Spectral evolution of the field launched into the fibre used in the experiment, for a 7-ps, 320-nJ pulse. The simulation is conducted using the generalized multimode nonlinear Schrödinger equation, with the first 10 modes of the fibre at 532 nm. The field is initially in the 8th mode, with a background of the other modes, each at 0.01% the energy launched into the 8th mode.

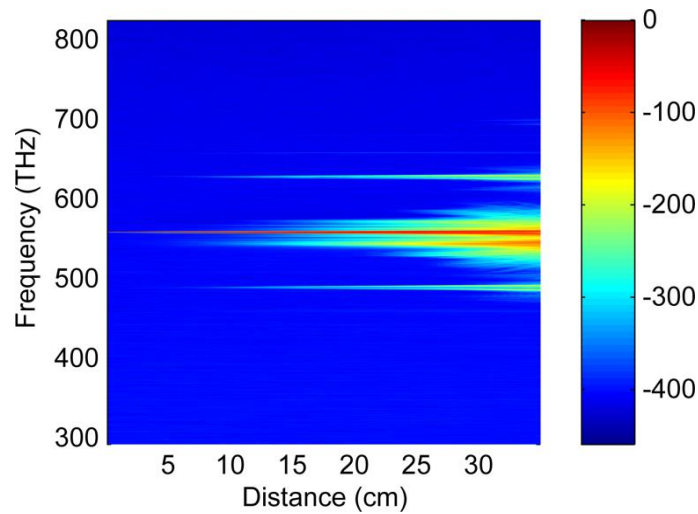

Supplementary Figure 18: Spectral evolution of the field launched into the fibre used in the experiment, for a 7-ps, 320-nJ pulse. The simulation is conducted using the generalized multimode nonlinear Schrödinger equation, with the first 10 modes of the fibre at 532 nm. The field is initially in the 1st and 4th modes, with a background of the other modes, each at 0.01% the energy launched into the 1st and 4th modes. The 1st and 4th modes are both radially-symmetric.

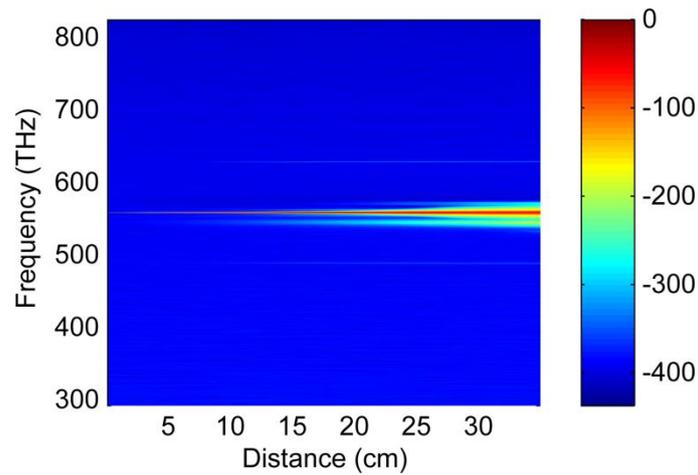

Supplementary Figure 19: Spectral evolution of the field launched into the fibre used in the experiment, for a 7-ps, 320-nJ pulse. The simulation is conducted using the generalized multimode nonlinear Schrödinger equation, with the first 10 modes of the fibre at 532 nm. The field is initially in all 10 modes equally.

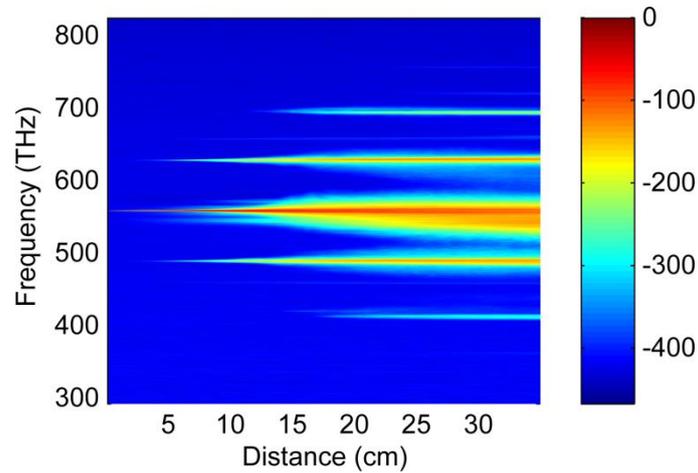

Supplementary Figure 20: Spectral evolution of the field launched into the fibre used in the experiment, for a 7-ps, 640-nJ pulse (note: twice the previous simulations). The simulation is conducted using the generalized multimode nonlinear Schrödinger equation, with the first 10 modes of the fibre at 532 nm. The field is initially in the 2nd and 6th modes, with a background of the other modes, each at 0.01% the energy launched into the 2nd and 6th modes. Even with double the energy as in Supplementary Figure 15, the growth of MI is still substantially smaller than when the attractor is launched.

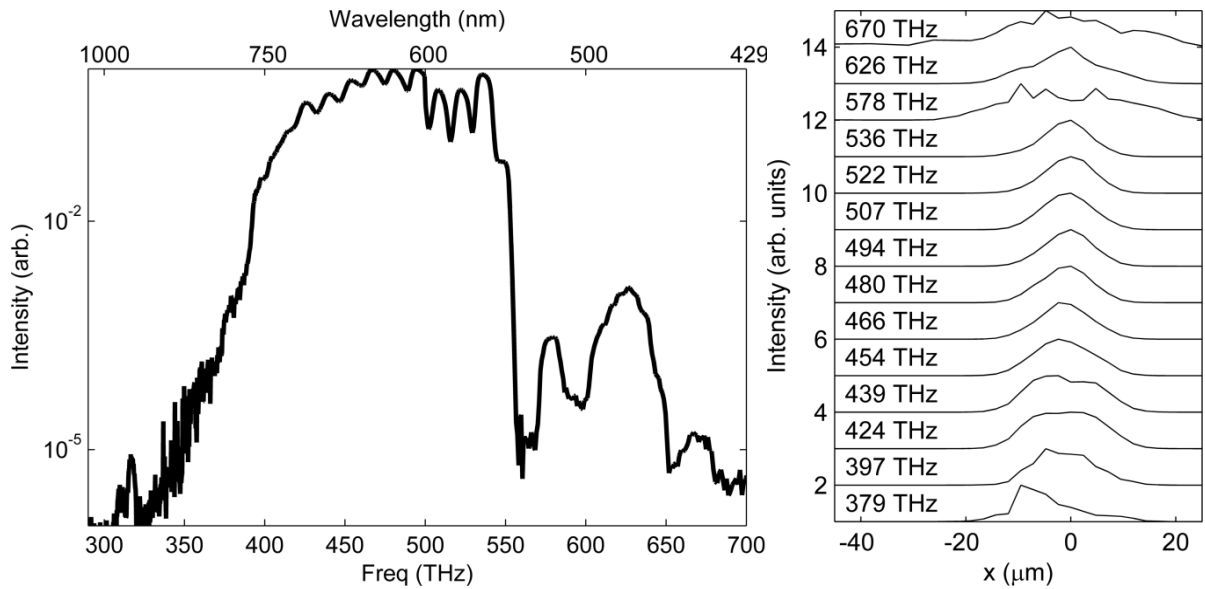

Supplementary Figure 21: An additional example of a supercontinuum spectrum and spatiospectral profiles for the field after instability has been excited. The fibre is the same one used for the experiments shown in Figure 1.

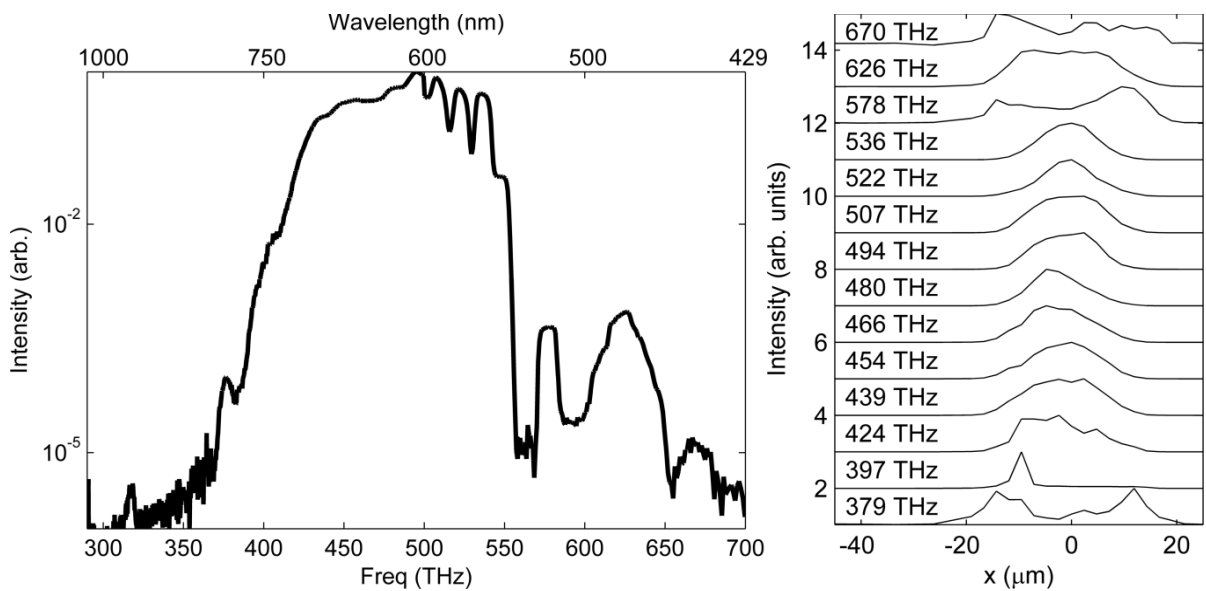

Supplementary Figure 22: An additional example of a supercontinuum spectrum and spatiospectral profiles for the field after instability has been excited. The fibre is the same one used for the experiments shown in Figure 1.

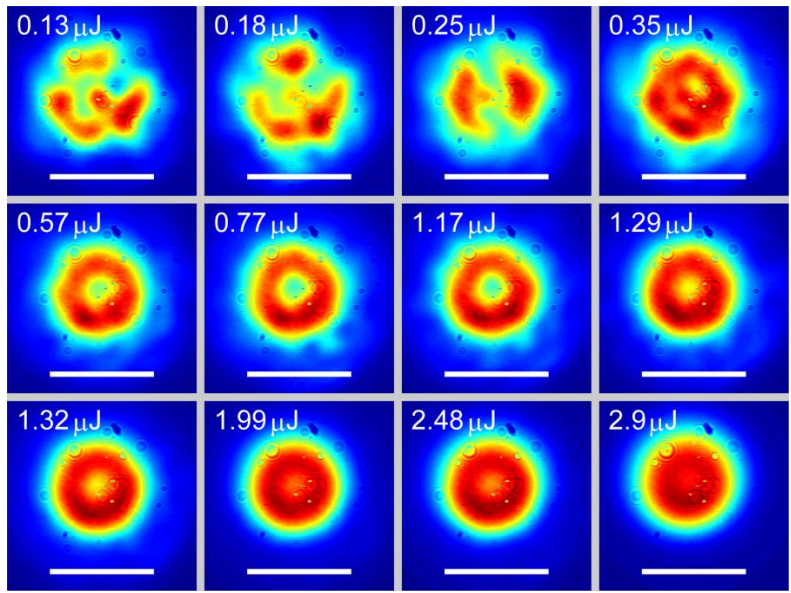

Supplementary Figure 23: near-field beam profile for increasing energy for an initial condition in which STMI is observed before the field completely reaches the attractor. Although the field does not completely reach the attractor, it does exhibit self-organized instability and appears to be approaching the attractor because the field becomes less-speckled and the lower-order mode content grows over the high-order mode content. Scale bar is 11 μm

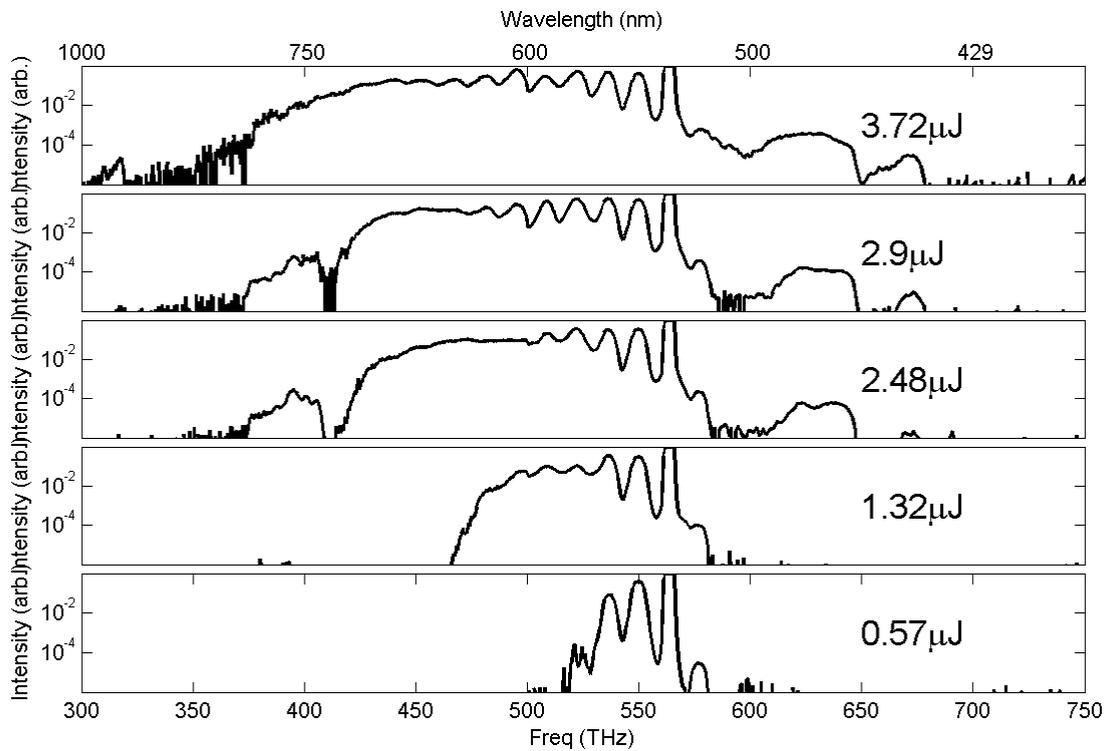

Supplementary Figure 24: Spectra for the same data set shown in Supplementary Figure 1.

To compare the nonlinearity in the experiment to that of a typical telecommunications transmission line, we compute the number of nonlinear lengths, $L_{nl}$, in the experiment. We then find the length of a typical telecommunications line with roughly the same number of nonlinear lengths. In terms of nonlinearity, these systems can then be considered equivalent since the field accumulates a similar total nonlinear phase ( $\varphi_{NL} = L/L_{nl}$ ).

For the telecommunications system, we consider the nonlinearity in a single mode (the full C-band of wavelength-division channels). For a multimode fibre transmission, the total power will be increased as each additional mode provides the same number of additional channels. Since the multimode fibre has a larger area, the intensity of the light ($I \propto A^{-1}$) in the fibre will be reduced. However, this is only in proportion with the growth of the number of spatial modes (the number of modes is $N \propto A / \lambda^2$ ). Since each spatial mode will increase the power launched into the fibre, the intensity, and therefore, the total nonlinearity of the transmission, is roughly the same as for a single spatial mode. Therefore, the nonlinearity in a single spatial mode transmission is a good estimate of the total nonlinearity in a multi-spatial-mode transmission.

The nonlinear length is $L_{nl} = 1/\gamma P_o$, where $P_o$ is the peak power and $\gamma$ is the nonlinear coefficient. $\gamma = \frac{2\pi n_2}{\lambda A_{eff}}$, where $n_2$ is the nonlinear index, $\lambda$ is the wavelength, and $A_{eff}$ is the effective mode area. For SMF, $A_{eff} = \pi R^2$ for $R$ the 1/e² diameter of the fundamental mode. For the GRIN fibre here, we take $R$ to be 0.85 times the core diameter, since the situation of interest is propagation of a multimode signal occupying roughly all modes equally.

To take loss into account, we compute the effective fibre length $L_{eff} = \frac{1-e^{-\alpha L}}{\alpha}$, where $\alpha$ is the loss of the fiber at the wavelength of interest. For telecommunications systems, $\lambda = 1550\ nm$ and for our experiment, $\lambda = 532\ nm$. In a transmission line, the loss is fully compensated by periodic or distributed amplifiers, so $\alpha \approx 0$. In our experiments, the loss is $\alpha \approx 10$ dB/km.

As is shown in the table below, we find the experimental conditions are similar, in terms of total nonlinearity, to a loss-managed telecommunications system of 150 km length.

**Supplementary Table 1: Calculation of nonlinearity for experiment and typical telecommunications system.**

|  | $P_o$ (W) | $n_2$ (m²/W) | $A_{eff}$ (μm²) | $\gamma$ (1/Wm) | $L_{nl}$ (m) | L (m) | $\alpha$ (1/m) | $L_{eff}$ (m) | $L_{eff}/L_{nl}$ |
|---|---|---|---|---|---|---|---|---|---|
| Current expt. | 500 | 2.3x10⁻²⁰ | 380 | 7.2x10⁻⁴ | 2.8 | 100 | 0.002304 | 89 | 32 |
| Typical SMF | 0.1 | 2.3 x10⁻²⁰ | 53 | 1.8x10⁻³ | 5600 | 150x10³ | ~0 | 150x10³ | 26 |